\documentclass[10pt]{iopart}
\usepackage{iopams}

\begin{document}

\title{Analytic approximations, perturbation methods,
and their applications}

\author{
P~Jaranowski$^1$\footnote{B3 session's chairman.
E-mail: pio@alpha.uwb.edu.pl.},
K~G~Arun$^{2,3,4}$,
L~Barack$^5$,
L~Blanchet$^3$,
A~Buonanno$^{6,7,3}$,
M~F~De~Laurentis$^{8,9}$,
S~Detweiler$^{10}$,
H~Dittus$^{11}$,
M~Favata$^{12}$,
G~Faye$^3$,
J~L~Friedman$^{13}$,
K~Ganz$^{14}$,
W~Hikida$^{15}$,
B~R~Iyer$^2$,
T~S~Keidl$^{13}$,
Dong-Hoon~Kim$^{16}$,
K~D~Kokkotas$^{17,18}$,
B~Kol$^{19}$,
A~S~Kubeka$^{20}$,
C~L\"ammerzahl$^{11}$,
J~Maj\'ar$^{21}$,
A~Nagar$^{8,9}$,
H~Nakano$^{22}$,
L~R~Price$^{10,13}$,
M~S~S~Qusailah$^{2,23}$,
N~Radicella$^{8,9}$,
N~Sago$^5$,
D~Singh$^{24}$,
H~Sotani$^{17}$,
T~Tanaka$^{14}$,
A~Tartaglia$^{8,9}$,
M~Vas\'uth$^{21}$,
I~Vega$^{10}$,
B~F~Whiting$^{10}$,
A~G~Wiseman$^{13}$,
S~Yoshida$^{25}$}

\address{$^1$ University of Bia{\l}ystok, Bia{\l}ystok, Poland}
\address{$^2$ Raman Research Institute, Bangalore, India}
\address{$^3$ Institut d'Astrophysique de Paris, CNRS, Paris, France}
\address{$^4$ Laboratoire de l'Acc\'el\'erateur Lin\'eaire, Orsay, France}
\address{$^5$ University of Southampton, Southampton, United Kingdom}
\address{$^6$ University of Maryland, College Park, Maryland, USA}
\address{$^7$ AstroParticule et Cosmologie, CNRS, Paris, France}
\address{$^8$ Politecnico di Torino, Torino, Italy}
\address{$^9$ INFN, Torino, Italy}
\address{$^{10}$ University of Florida, Gainesville, Florida, USA}
\address{$^{11}$ University of Bremen, Germany}
\address{$^{12}$ Kavli Institute for Theoretical Physics, Santa Barbara, California, USA}
\address{$^{13}$ University of Wisconsin-Milwaukee, Milwaukee, Wisconsin, USA}
\address{$^{14}$ Kyoto University, Kyoto, Japan}
\address{$^{15}$ Osaka University, Toyonaka Osaka, Japan}
\address{$^{16}$ Albert Einstein Institute, Potsdam, Germany}
\address{$^{17}$ University of Thessaloniki, Thessaloniki, Greece}
\address{$^{18}$ University of T\"ubingen, T\"ubingen, Germany}
\address{$^{19}$ Hebrew University, Jerusalem, Israel}
\address{$^{20}$ University of South Africa, South Africa}
\address{$^{21}$ Research Institute for Particle and Nuclear Physics, Budapest, Hungary}
\address{$^{22}$ University of Texas at Brownsville, Brownsville, Texas, USA}
\address{$^{23}$ Sana'a University, Yemen}
\address{$^{24}$ University of Regina, Regina, SK, Canada}
\address{$^{25}$ University of Tohoku, Sendai, Japan}

\begin{abstract}

The paper summarizes the parallel session B3
{\em Analytic approximations, perturbation methods, and their applications}
of the GR18 conference.
The talks in the session reported notably recent
advances in black hole perturbations and post-Newtonian approximations
as applied to sources of gravitational waves.

\end{abstract}


\submitto{\CQG}


\section{Introduction}

Analytic approximations and perturbation methods
are of common use in different branches of general relativity theory.
The most common methods are the post-Newtonian approximations,
able to deal with the gravitational field of any system
in the non-relativistic limit,
the post-Minkowskian approximations,
which are appropriate for relativistic systems
in the weak gravitational field regime,
and the perturbation formalisms, which expand
(at linear order in general)
around some exact solution of the Einstein field equations.
The B3 session {\em Analytic approximations,
perturbation methods, and their applications}
of GR18 conference was dominated by the issues
more or less directly related to the problem of detecting gravitational waves
by currently operating or planned to be built in the near future detectors.
One of the most important targets for gravitational-wave detectors
are coalescing binary systems made of compact objects of different kinds.
These sources produce ``chirps'' of gravitational radiation
whose amplitude and frequency are increasing in time.
For successful detection of the chirp signals
and extraction of their astrophysically important parameters
it is crucial to theoretically predict
gravitational waveforms with sufficient accuracy.

One of the important sources for the future space-based LISA detector
are extreme-mass-ratio binaries
consisting of a small compact body (of stellar mass)
orbiting around a supermassive black hole.
The need for accurate modelling of the orbital dynamics of such binaries
motivates some recent work on the problem of calculating
the gravitational self-force experienced by a point particle moving
in the background space-time of a more massive body.
Several talks in the session reported the progress
on different aspects of the gravitational self-force computations.
These were talks by {\em S~Detweiler}, {\em M~Favata},
{\em J~L~Friedman}, {\em W~Hikida}, {\em N~Sago}, and {\em B~F~Whiting}.

For comparable-mass binaries,
computations based on the post-Newtonian approximation of general relativity
are useful for constructing gravitational-wave templates
for data-analysis purposes.
The post-Newtonian approximation describes
with great accuracy the inspiral phase of these systems.
The analytic post-Newtonian results are currently matched
to recent numerical calculations of the merger and ring-down phases
of black hole binaries \cite{Buonanno&Cook&Pretorius2007,Boyle&al2007}.
In the contributed talks
(by {\em L~Blanchet}, {\em B~R~Iyer}, and {\em M~Vas\'uth})
recent results on incorporating the spin-orbit effects
(also within the accuracy beyond the leading-order effects)
as well as the generalization to eccentric orbits
(most of the explicit analytic results concern circular orbits)
were presented.

Recently a new approach to the perturbative solution
of the problem of motion and radiation in general relativity was developed.
This is the approach pioneered
by Goldberger and Rothstein \cite{Goldberger&Rothstein2006}
in which effective field theory methods
were used to describe the dynamics
of nonrelativistic extended objects coupled to gravity.
{\em B~Kol} discussed this approach
as well as some of its applications.
{\em A~Tartaglia} presented a new semi-analytic method
for computing the emission of gravitational waves
by a compact object moving
in the background of a more massive body.

Several other topics were tackled in the contributed talks.
{\em C~L\"ammerzahl} discussed the influence of the cosmic expansion
on the physics in gravitationally bound systems.
{\em D~Singh} presented an analytic perturbation approach
for classical spinning particle dynamics,
based on the Mathisson-Papapetrou-Dixon equations of motion.
{\em H~Sotani} studied gravitational radiation
from collapsing magnetized dust.
{\em A~S~Kubeka} remarked on the computation
of the Ricci tensor for non-stationary axisymmetric space-times.

In the rest of this article
all talks contributed to the B3 session are sketched in more detail
(in the order in which they were presented at the conference).

\section{Contributed talks}

\subsection{{\em Self-force analysis in extreme-mass-ratio inspiral}
by S~Detweiler and I~Vega (reported by S~Detweiler)}

The motion of a small object of mass $m$ orbiting a supermassive black
hole of mass $M$ deviates slightly from a geodesic and has an acceleration
that scales as the ratio $m/M$ of the masses. This acceleration includes
the dissipative effects of radiation reaction and is said to result from
the gravitational self-force acting on $m$ \cite{Poisson2004}. As an
alternative, the effects of the self-force may be described as geodesic
motion in an appropriately regularized metric of the perturbed spacetime
\cite{Detweiler2005}.
The LISA effort requires accurate gravitational wave templates for data analysis.
For extreme-mass-ratio inspirals the templates should include
both the dissipative and conservative effects of the self-force.

The talk described a novel, efficient method
for simultaneously calculating both the gravitational self-force
as well as its effect on the gravitational waveform.
The Authors replaced the usual singular point source
with a distributed, abstract, analytically determined source.
The resulting perturbation in the field from this special distributed source
is guaranteed to be differentiable at the location of the particle
and to provide the appropriate self-force effect on the motion of the particle.
At the same time, the field from the distributed source
is identically equal to the actual perturbed field in the wave zone.
Thus, this abstract field simultaneously provides
both the self-force acting on a point source
and also the effect of the self-force on the waveform of the radiation.

\subsection{{\em The adiabatic approximation
and three-body effects in extreme-mass-ratio inspirals} by M~Favata}

Extreme-mass-ratio inspirals (EMRIs) are an important class of LISA sources
consisting of a compact object inspiralling into a supermassive black hole.
The detection of these sources and the precision measurement
of their parameters relies on the accurate modeling of their orbital dynamics.
A precise description of the binary's orbit requires
an evaluation of the compact object's self-force.
The adiabatic approximation (more appropriately
referred to as the {\em radiative approximation}
\cite{Pound&Poisson2007a,Pound&Poisson2007b}),
consists of computing the time-averaged rates of change
of the three conserved quantities of geodesic motion.
Its use greatly simplifies the computation of the orbital evolution.
However, the adiabatic approximation ignores corrections
to the conservative dynamics proportional
to the mass $m$ of the compact object.
These `post-adiabatic' corrections
will affect the binary's positional orbital elements,
for example, by causing $O(m)$ corrections to the pericenter precession rate.

Using a toy model
of an electric charge orbiting a central mass
and perturbed by the electromagnetic self-force,
Pound, Poisson, and Nickel \cite{Pound&Poisson&Nickel2005}
have called into question
the accuracy of the adiabatic approximation,
especially for eccentric orbits.
In order to estimate
the size of the post-adiabatic phase errors
in the gravitational case,
the Author presented an analytical computation,
accurate to second-post-Newtonian order,
of the small-eccentricity corrections
to the gravitational-wave phase.
These post-Newtonian, eccentricity corrections
to the phase can be significant
not only for EMRIs but for other binary sources as well.
Using this phase expansion it was found
that the post-adiabatic phase errors
are sufficiently small that waveforms
based on the adiabatic approximation
can be used for EMRI detection,
but not for precise parameter measurements.

The Author also discussed
the effect of a third mass orbiting the EMRI.
The analysis models the system as a hierarchical triple
using the equations of motion
of Blaes, Lee, and Socrates \cite{Blaes&Lee&Socrates2002}.
To have a measurable effect on the EMRI's waveform,
the distant mass must be sufficiently close to the compact object
that both the inner and outer binaries would be detected as EMRIs.
Such ``double-EMRI'' systems are rare.

\subsection{{\em Extreme-mass-ratio binary inspiral in a radiation gauge}
by~J~L~Friedman, T~S~Keidl, Dong-Hoon~Kim, E~Messaritaki and A~G~Wiseman
(reported by~J~L~Friedman)}

Gravitational waves from the inspiral of a stellar-size black hole of mass $m$
to a supermassive black hole of mass $M$ can be accurately approximated
by a point particle moving in a Kerr background.
To find the particle's orbit to first order in the mass ratio $m/M$,
one must compute the self-force. The computation requires a renormalization,
but the well-known MiSaTaQuWa prescription
\cite{Mino&Sasaki&Tanaka1997,Quinn&Wald1997} involves a harmonic gauge,
a gauge that is not well suited to perturbations of the Kerr geometry.
In a harmonic gauge, one solves ten coupled PDEs,
instead of the single decoupled Teukolsky equation
for the gauge invariant components ($\psi_0$ or $\psi_4$) of the perturbed Weyl tensor.
In the talk progress was reported
in finding the renormalized self-force from $\psi_0$ or $\psi_4$.
Following earlier work by Chrzanowski and by Cohen and Kegeles,
a radiation gauge was adopted
to reconstruct the perturbed metric from the perturbed Weyl tensor.
The Weyl tensor component is renormalized by subtracting a singular part
obtained using a recent Detweiler-Whiting version \cite{Detweiler&Whiting2003}
of the singular part of the perturbed metric
as a local solution to the perturbed Einstein equations.
The Authors' method relies on the fact that the corresponding renormalized $\psi_0$
is a {\em sourcefree} solution to the Teukolsky equation.
One can then reconstruct a nonsingular renormalized metric in a radiation gauge,
a gauge that exists only for vacuum perturbations.
More details can be found in Ref.\ \cite{Keidl&al2007}.

\subsection{{\em Adiabatic evolution of three `constants' of motion in Kerr spacetime}
by~W~Hikida, K~Ganz, H~Nakano, N~Sago and T~Tanaka (reported by~W~Hikida)}

General orbits of a particle of small mass around a Kerr black hole
are characterized by three parameters:
the energy, the angular momentum, and the Carter constant.
For energy and angular momentum, one can evaluate their change rates
from the fluxes of the energy and the angular momentum at infinity
and on the event horizon according to the balance argument.
On the other hand, for the Carter constant,
one can not use the balance argument
because the conserved current associated with it is not known.
Recently Mino proposed a new method
of evaluating the averaged change rate of the Carter constant
by using the radiative field.
The Authors developed a simplified scheme for practical evaluation
of the evolution of the Carter constant based on the Mino's proposal.
In the talk this scheme was described in some detail,
and derivation of explicit analytic formulae for the change rates
of the energy, the angular momentum, and the Carter constant was presented.
Also some numerical results for large eccentric orbits were shown.
For more details see Ref.\ \cite{Ganz&al2007}.

\subsection{{\em Gravitational self-force
on a particle orbiting a Schwarzschild black hole}
by~L~Barack and N~Sago (reported by~N~Sago)}

In the talk the calculation of the gravitational self-force
acting on a pointlike particle moving around a Schwarzschild black hole
was presented. The calculation was done in the Lorenz gauge.
First, the Lorenz-gauge metric perturbation equations
were solved directly using numerical evolution in the time domain.
Then the back-reaction force
from each of the multipole modes of the perturbation was computed.
Finally, the {\em mode sum} scheme was applied
to obtain the physical self-force.
The temporal component of the self-force
describes the rate of the loss of orbital energy.
As a check of their scheme,
the Authors compared their result for this component
with the total flux of gravitational-wave energy
radiated to infinity and through the event horizon.
The radial component of the self-force was also calculated.
From their result for the self-force,
the Authors computed the correction to the orbital frequency
due to the gravitational self-force,
taking into account of both the dissipative and the conservative effects.
More details can be found in Ref.\ \cite{Barack&Sago2007}.

\subsection{{\em Mobile quadrupole as a semianalytic method for gravitational-wave emission}
by~A~Tartaglia, M~F~De~Laurentis, A~Nagar and N~Radicella (reported by~A~Tartaglia)}

The quadrupole formula is the simplest approximation
for studying the gravitational-wave emission from a binary system.
The formula gives its best performance
for quasi-circular and quasi-stationary motion of the emitters.
Whenever the motion is far from the quasi-circular approximation,
other semi-analytic methods
or numerical calculations of growing complexity must be used.
In the talk a situation was studied
where the gravitational wave is emitted by a concentrated object
accelerating in the background field of a central mass.
Provided one knows the background metric,
the gravitational wave represents a first order perturbation on it,
so that the space-time trajectory of each object of the pair is almost geodesic.
Once the equations of a geodesic are written,
the motion can be thought of as an instantaneous rotation
around an (instantaneously at rest) curvature centre for the space trajectory.
In this condition the quadrupole formula is easily applicable
at each different place along the geodesic,
after calculating the curvature and the equivalent angular velocity
as the ratio between the three-speed and the curvature radius.
Everything is reduced to a problem of ordinary geometry.
The energy emission rate and the waveforms obtained by this way
must simply be converted from the local time
to the time of a far away inertial observer.
The approach was applied
to the capture of a mass by a Kerr black hole.
The method is much lighter
(from the computational point of view)
than numerical relativity, giving comparable results.
For the research results relevant to the presented approach see Refs.\
\cite{Dymnikova1977,Dymnikova&Popov1980,Schnittma&Bertschinger2004}.

\subsection{{\em The non-radiated multipoles in the perturbed Kerr spacetime}
by~L~R~Price and B~F~Whiting (reported by~B~F~Whiting)}

For the self-force problem in general relativity,
it has been shown that the perturbed metric
produced by a finite-mass test, point-particle,
has a singular part which exerts no influence on the particle,
while the self-force which the particle experiences
arises entirely due to a metric perturbation
which is smooth at the location of the particle \cite{Detweiler&Whiting2003}.
However, metric reconstruction from the perturbed Weyl tensor
is unable to yield perturbations
for the non-radiated multipoles in Petrov type II spacetimes,
such as that surrounding the Kerr black hole \cite{Whiting&Price2005}.
In the talk a new form of the perturbed Einstein equations,
developed by the Authors on the base of the Newman-Penrose formalism,
was presented. With its assistance,
progress towards filling the low multipole gap,
which will contribute to the calculation
of regularization parameters for the self-force problem, was discussed.

\subsection{{\em Higher-order spin effects in the radiation field of compact binaries}
by~L~Blanchet, A~Buonanno and G~Faye (reported by~L~Blanchet)}

In the talk the investigation, motivated by the search for
gravitational waves emitted by binary black holes, of the gravitational
radiation field of compact objects with spins was discussed. The
approach is based on the multipolar post-Newtonian wave generation
formalism and on the formalism of point particles with spin
(Papapetrou-Dixon-Bailey-Israel). The Authors computed: (i) the
spin-orbit coupling effect in the binary's equations of motion one
post-Newtonian (PN) order beyond the dominant effect (confirming a
previous result by Tagoshi et al.\ \cite{Tagoshi&Ohashi&Owen2001}),
(ii) the spin-orbit coupling effects
in the binary's mass and current quadrupole moments at the same order,
(iii) the spin-orbit contributions in the gravitational-wave energy flux,
and (iv) the secular evolution of the binary's orbital phase up to 2.5PN order
(for maximally rotating black holes).
Previous results on the spin-orbit effect at the lowest order
were computed in Refs.\ \cite{Kidder&Will&Wiseman1993,Kidder1995}.
Crucial ingredients for obtaining the next-order 2.5PN contribution in the
orbital phase are the binary's energy and the spin precession equations.
These results provide more accurate gravitational-wave templates to be
used in the data analysis of rapidly rotating Kerr-type black-hole
binaries with the ground-based interfrometric detectors and the
space-based detector LISA.
Details of the presented results were published in Refs.\
\cite{Faye&Blanchet&Buonanno2006,Blanchet&Buonanno&Faye2006}.

\subsection{{\em The 3PN gravitational wave luminosity
from inspiralling compact binaries in~eccentric orbits}
by~K~G~Arun, L~Blanchet, B~R~Iyer and M~S~S~Qusailah (reported by~B~R~Iyer)}

Some details of the computation
of the complete gravitational-wave luminosity of inspiralling compact binaries
on quasi-elliptical orbits up to the third post-Newtonian (3PN) order
using multipolar post-Minkowskian formalism were presented.
There are two types of contributions
to the gravitational-wave luminosity at 3PN order:
the instantaneous type terms,
which depend on the dynamics of the binary only at the retarded instant,
and the hereditary terms,
which are sensitive to dynamics of the system in the entire past.
The new inputs for the calculation of the 3PN instantaneous terms
include the mass octupole and current quadrupole at 2PN for general orbits
and the 3PN accurate mass quadrupole.
Using the 3PN quasi-Keplerian representation of elliptical orbits
obtained recently \cite{Memmesheimer&Gopakumar&Schafer2004},
the flux is averaged over the binary's orbit.
The hereditary terms have a `tail', `tail of tail' and `tail-squared' contributions
which are computed using a semi-analytic procedure
extending the earlier work of Blanchet and Sch\"afer at 1.5PN
\cite{Blanchet&Schafer1993}.
This semi-analytic extension uses
the 1PN quasi-Keplerian parametrisation of the binary
and exploits the doubly periodic nature of the orbital motion.
The final 3PN accurate energy flux averaged over the binary's orbit was presented
in the modified harmonic (which contains no logarithmic terms) and the ADM coordinates.
Also a gauge invariant expression of the flux was provided
in terms of the orbital frequency and the periastron precession constant.
The results are consistent with those obtained by perturbation theory
in the test particle limit to order $e_t^2$
(where $e_t$ is the so-called time eccentricity)
and the 3PN circular orbit results.
These results form the starting input
for the construction of templates for inspiralling binaries in quasi-eccentric orbits,
an astrophysically possible class of sources
both for the ground-based and the space-based gravitational-wave interferometers.

\subsection{{\em On the influence of the cosmic expansion
on the physics in gravitationally bound systems}
by~C~L\"ammerzahl and H~Dittus (reported by~C~L\"ammerzahl)}

It is an old question whether the cosmological expansion influences
the dynamics of gravitationally bound systems
\cite{Lammerzahl&Preuss&Dittus2007}.
Though sometimes it has been claimed
that the expansion will tie apart gravitationally bound systems,
the majority of papers covering this issue derive no measurable influence.
In the talk some additional arguments for the latter were given.
It was shown that (i) the gravitational field created by an isolated body
will feel only a tiny influence,
(ii) the planetary orbits also are practically inert against the expansion,
and (iii) Doppler tracking of satellites in deep space
is also only marginally influenced by the cosmic expansion.

\subsection{{\em Spin evolution in binary systems}
by~M~Vas\'uth and J~Maj\'ar (reported~by~M~Vas\'uth)}

Gravitational waves emitted by compact binary systems are characterized
by different parameters of the binary. Among them the effects of rotation
of the orbiting bodies appear at 1.5 post-Newtonian (PN) order both in the
dynamical description and the wave generation problem. In the talk the
evolution of the individual spins of the bodies was discussed for compact
binaries in circular and general eccentric orbits.
For a 2PN description the spin precession equations
were analyzed up to 0.5PN order.
To the lowest order
the angles between the total angular momentum and spin vectors
are constant and spin-spin interaction
causes additional harmonic dependence.
The true anomaly parameterization
proved to be useful in the description of eccentric orbits.
In the circular and general cases
linear and harmonic dependence of the angles
describing the orientation of spins were found.

\subsection{{\em Matched asymptotic expansion
as a classical effective field theory} by~B~Kol}

In the talk it was explained
how the method of {\em classical effective field theory}
borrowed from {\em quantum field theory}
by Goldberger and Rothstein \cite{Goldberger&Rothstein2006}
in the context of the motion of a compact object
within a background whose typical length scale is much larger,
is equivalent to {\em matched asymptotic expansion},
and moreover it offers additional insight.
Feynman diagrams, divergences, (dimensional) regularization, counter-terms
and the Feynman gauge appeared.
The ideas were demonstrated by the case of caged black holes
(black holes within a compact dimension).
Within this method the source
is replaced by a "black box" effective action.
Another application of these ideas is to the inspiral problem of a binary system.
The Author presented a computation utilizing high energy physics methods
of the radiation reaction force for the case of a scalar field.

\subsection{{\em An analytic perturbation approach
for classical spinning particle dynamics} by~D~Singh}

The Author presented a perturbation method
to analytically describe the dynamics of a classical spinning particle,
based on the Mathisson-Papapetrou-Dixon equations of motion.
By a power series expansion with respect to the particle's spin magnitude,
it was demonstrated how to obtain an analytic representation
of the particle's kinematic and dynamical degrees of freedom
that is formally applicable to infinite order in the expansion parameter.
Within this formalism, it is possible to identify
a classical analogue of radiative corrections
to the particle's mass and spin due to spin-gravity interaction.
The robustness of this approach
was demonstrated by showing how to explicitly compute
the first-order momentum and spin tensor components
for arbitrary particle motion in a general space-time background.
Potentially interesting applications
based on this perturbation approach were also considered.
For more details see Ref.\ \cite{Singh2007}.

\subsection{{\em Gravitational radiation from collapsing magnetized dust}
by~H~Sotani, S~Yoshida and K~D~Kokkotas (reported by~H~Sotani)}

The Authors studied the influence of magnetic fields
on the axial gravitational waves emitted during
the collapse of a homogeneous dust sphere.
It was found that while the energy emitted
depends weakly on the initial matter perturbations
it has strong dependence on the strength and the distribution
of the magnetic field perturbations.
The gravitational wave output of such a collapse
can be up to an order of magnitude larger or smaller
calling for detailed numerical 3D studies of collapsing magnetized configurations.
More details are given in Ref.\ \cite{Sotani&al2007}.

\subsection{{\em Gravitational waveforms for compact binaries}
by~M~Vas\'uth and J~Maj\'ar (reported by~M~Vas\'uth)}

Among the promising sources of gravitational radiation
are binary systems of compact stars.
The detectable signal is characterized by different parameters of the system,
e.g., rotation of the bodies and the eccentricity of the orbit.
The Authors presented a method
to evaluate the gravitational wave polarization states
for inspiralling compact binaries
and considered eccentric orbits
and the spin-orbit contribution
in the case of two spinning objects up to 1.5 post-Newtonian order.
In the circular orbit limit the presented results
are in agreement with existing results.
For more details see Ref.\ \cite{Vasuth&Majar2007}.

\subsection{{\em On the Ricci tensor for non-stationary axisymmetric space-times}
by~A~S~Kubeka}

The results on Ricci tensor for non-stationary axisymmetric space-times
determined by Chandrasekhar \cite{Chandrasekhar1975}
have been found to be incorrect both in the linear and non-linear regimes.
However, the incorrectness of the Ricci tensor
does not affect the well-known results on linear perturbations of a Schwarzschild
black hole solution.

\section*{Acknowledgments}

MV contribution was supported by OTKA grant No.\ F049429.
The work of SD, LP, IV and BW was supported by NSF Grant No.\ PHY-0555484.

\end{document}